\documentclass[a4paper,twocolumn,superscriptaddress]{revtex4-1}
\usepackage{amsmath}
\usepackage{graphicx}
\newcommand{\nn}{\nonumber\\}

\begin{document}

\title{Atomistic theory of electronic and optical properties of InAsP/InP nanowire quantum dots }
\author{Moritz Cygorek}
\affiliation{Department of Physics, University of Ottawa, Ottawa, Canada K1N 6N5}
\author{Marek Korkusinski}
\affiliation{Department of Physics, University of Ottawa, Ottawa, Canada K1N 6N5}
\affiliation{Security and Disruptive Technologies,
National Research Council of Canada, Ottawa, Canada K1A OR6}
\author{Pawel Hawrylak}
\affiliation{Department of Physics, University of Ottawa, Ottawa, Canada K1N 6N5}

\begin{abstract}
We present here an atomistic theory of the electronic and optical properties of hexagonal 
InAsP quantum dots in InP nanowires in the wurtzite phase.  These self-assembled 
quantum dots are unique in that their heights, shapes, and diameters are
well known. Using a combined valence-force-field, tight-binding, and configuration-interaction
approach we perform atomistic calculations of single-particle states 
and excitonic, biexcitonic and trion complexes as well as emission spectra 
as a function of the quantum dot height, diameter and As versus P concentration.
The atomistic tight-binding parameters
for InAs and InP in the wurtzite crystal 
phase were obtained by \emph{ab initio} methods corrected by empirical band gaps.  
The low energy electron and hole states form electronic shells 
similar to parabolic or cylindrical quantum confinement,
only weakly affected by hexagonal symmetry and As fluctuations. 
The relative alignment of the emission lines from excitons, trions and 
biexcitons agrees with that for InAs/InP dots in the 
zincblende phase in that biexcitons and positive trions are
only weakly bound. 
The random distribution of As atoms leads
to dot-to-dot fluctuations of a few meV for the single-particle states and the 
spectral lines. Due to the high symmetry of hexagonal InAsP nanowire
quantum dots the exciton fine structure splitting is found to be small,  
of the order a few $\mu$eV with significant random fluctuations 
in accordance with experiments.
\end{abstract}

\maketitle

\section{Introduction}
Semiconductor quantum dots \cite{book_SQD, book_QD}
are promising structures for quantum devices for quantum technologies, 
in particular due to their interaction with light. This includes
quantum dot lasers\cite{Arakawa_QD}, 
single-photon sources\cite{singlephoton_Michler,singlephoton_Cosacchi,
singlephoton_Santori,Review_Mantynen}, emitters of entangled photon pairs
\cite{entangled_Orieux,entangled_Stevenson2006,Dalacu_entangled_2014,
Dalacu_entangled_2019,concurrence_Cygorek, concurrence_Tim, 
Korkusinski_photon_cascades} and highly entangled photon cluster states
\cite{clusterstate_proposal,clusterstate_measured} 
and other quantum information applications\cite{Reimer_scalable, Cogan2018, synthetic_haldane_chain}.
One major obstacle for the  self-assembled semiconductor quantum-dot-based 
devices is the variation of size, shape, and position of
quantum dots\cite{selfassembled,selfassembled_Wang}.
This implies that also their
electronic and optical properties vary from dot to dot and the 
necessary selection of dots with the desired characteristics leads to a 
very small yield of the overall device fabrication process.

One approach to deterministically build quantum dots with specific 
geometries is the growth of quantum dots in nanowires, 
where the shape and diameter of the quantum dot is defined by the shape and 
diameter of the nanowire,
 and the height as well as the material composition of the dot can be 
controlled during the vertical growth of the nanowire.
In Ref. \onlinecite{Dalacu_pureWZ} a selective-area vapour-liquid-solid (VLS) 
growth technique was employed to fabricate hexagonal InAs$_{x}$P$_{1-x}$ 
quantum dots in InP nanowires. 
The growth of nanowires with small diameters has been shown to produce
pristine crystals in the wurtzite phase with negligible intermixing of
zincblende stacking \cite{Shtrikman_stacking_faults}.
The high symmetry of the hexagonal nanowire quantum dots results 
in a strongly reduced 
exciton fine structure splitting\cite{Bester_smallFSS,Lixin_He_FSS155, Kadantsev_fine_structure, Dalacu_entangled_2014}. 
As a consequence a very high degree of 
entanglement of photon pairs emitted from nanowire quantum dots via the
biexciton cascade has been demonstrated\cite{Dalacu_entangled_2014,
Dalacu_entangled_2019}.
Furthermore, it is also possible to deterministically grow multiple dots 
within the same nanowire\cite{Hughes_doubledot}, which allows studying 
coherent coupling between dots\cite{Bayer_Science_molecules}.

Because of the recent progress in the fabrication of and experiments on InAsP 
nanowire quantum dots (for a review cf. Refs.~\onlinecite{Dalacu_review,
Review_Mantynen})
it is highly desirable to be able to understand and atomistically simulate 
their electronic and optical properties.
This is, however, a challenging task for a number of reasons: 

First, the number of atoms in a quantum dot and its immediate surrounding in
the nanowire approaches hundreds of thousands to millions of atoms.
Second, the nanowires and dots typically grow
in wurtzite crystal structure, whereas bulk InP and InAs form zincblende
lattices, hence information about bulk wurtzite materials is scarce.
Third, the As atoms in InAs$_{x}$P$_{1-x}$ dots are randomly incorporated, 
which gives rise to strong local fluctuations of the confining potential.
Fourth, the lattice mismatch between InAs and InP is about
$3\%$ so that strain in the wurtzite structure has to be taken into account. The random incorporation
of As atoms will also lead to strong spatial fluctuations of the strain
field. Thus, strain should be accounted for atomistically as well.

Some of these challenges have been addressed for self-assembled InAs/InP 
quantum dots with zincblende structure, which have been studied experimentally 
\cite{Dalacu_directed_selfassembly,Noetzel_anisotropy} 
and theoretically \cite{Zielinski_selfassembled_InAs, 
Weidong_selfassembled_InAs,Jaskolski_strain_effects,
Lixin_He_psp, Lixin_He_FSS155} 
by a number of groups, including some of us.
Here, however, we are interested in InAs quantum dots in InP nanowires,
where the main challenge remains the wurtzite structure of InAs$_{x}$P$_{1-x}$
nanowire quantum dots and the scarcity of empirical data.
For example, no empirical band structures are available, 
which are often the starting point of the theoretical description of 
nanostructures.
Furthermore, wurtzite and zincblende phases give rise to qualitatively 
different band structures with a relevant crystal field splitting
in the wurtzite phase and the difference in band gaps between, e.g., wurtzite
and zincblende InP of about 80 meV is large enough to enable the formation of 
polytypic zincblende-wurtzite quantum dots or wells in pure InP 
nanowires\cite{InP_polytypic_NW}. It is therefore difficult to infer concrete
information on wurtzite structures from data on the zincblende phase.
For these reasons one has to resort mostly to \emph{ab initio} methods.

So far, $k\cdot p$ theory parameters for InP and InAs in the wurtzite phase 
have been obtained from DFT calculations in Ref.~\onlinecite{wurtzite_KP}.
On the other hand, atomistic tight-binding calculations of InAs/InP nanowire
quantum dots\cite{Niquet_InAsP,Zielinski_elongated2013,Zielinski_light_hole2013}
and quantum dot molecules~\cite{Zielinski_molecules} have been performed,
but using parameters for the zincblende crystal phase \cite{Jancu1998}. 

In this article, we develop a description of InAs$_{x}$P$_{1-x}$
nanowire quantum dots in the wurtzite phase based on valence-force-field, 
tight-binding, and configuration-interaction methods.
The procedure of the simulations follows that of the computational toolkit
QNANO described in Ref.~\onlinecite{Zielinski_selfassembled_InAs,Marek_review}. However, for the purpose
of this project we have completely refactored the QNANO toolkit 
and the core elements have been parallelized to run efficiently on computer
clusters with tens of thousands of cores enabling simulations involving millions of atoms.

Central for modelling the nanowire quantum dots are the 
tight-binding parameters, including strain corrections.
For InAs and InP, these parameters have been available only for the zincblende crystal phase
\cite{Jancu1998}. Here, we obtain a set of parameters for
the wurtzite phase by fits to DFT band structures corrected 
to reproduce experimentally known band gaps.
Using those parameters, we then present simulations of single-particle states 
and excitonic complexes for typical hexagonal InAsP nanowire quantum dots 
and study their dependence on the dot height, diameter, As concentration and profile 
as well as the effects of the intrinsic randomness of As alloying in the InP 
matrix.

We find that the single-particle states, in particular the conduction band 
states, form shells that can be classified according to their angular momentum, 
e.~g., s-, p-, or d-type states, in analogy to lens shaped cylindrical quantum
dots\cite{Hawrylak_shells}. The high symmetry of the quantum dot is reflected in the high symmetry of 
the charge densities of the single-particle states. 
Increasing the quantum dot height or 
diameter reduces the confinement energy and decreases the single particle gap.
An increased As concentration results in deeper confinement potentials, 
which also decreases the single-particle gap. Furthermore, investigations
of the consequences of a delayed incorporation of As atoms, which leads to
As concentration gradients along the growth direction, show that the 
gap increases with increasing delay lengths. 
Calculations of the lowest-energy exciton, biexciton and trion states
predict a characteristic  alignment of spectral lines, where the biexciton 
emission line is below the exciton emission line, the line originating from 
negatively charged trion 
lies energetically below the biexciton line and the positively charged trion
is located close to the bright exciton line. The fine structure splitting of
the bright exciton states, which is technologically very important, e.~g.,
for the degree of entanglement of emitted photon pairs in the biexciton 
cascade, is predicted to be about 7 $\mu$eV for InAs$_{0.2}$P$_{0.8}$ 
dots with a height of 4 nm and a diameter of 18 nm.

The article is structured as follows: First, we briefly review the 
valence-force-field, tight-binding and configuration-interaction approach
used for the simulation of quantum dots and we provide details about their 
numerical implementation. Then, we describe the band structure
calculations using DFT, the correction of the band gaps and additional 
quantities entering the modelling of strained bulk wurtzite InAs and InP. 
This is followed by a description of 
the fitting procedure and the tight-binding parameters for 
InAs and InP in the wurtzite crystal phase. Finally, we present 
and summarize the results of the simulations for quantum dots.

\section{Methods}
To simulate the electronic and optical properties of quantum nanostructures, 
we employ a combined
valence-force-field, tight-binding, and configuration-interaction approach
reviewed in more detail in Refs.~\onlinecite{Zielinski_selfassembled_InAs,Marek_review}. Here, we
only present a brief overview.

\subsection{Strain relaxation}
The starting point for the simulations are the approximate positions 
and elements of every atom in the semiconductor nanostructure. 
Because there is a significant lattice mismatch of
about 3\% between InAs and InP, strain effects are important. 
Therefore, in a first step, 
the atomic positions are relaxed by minimizing the total elastic energy
according to the valence force field (VFF) method\cite{Keating}
\begin{align}
U&=\frac 12\sum_{i=1}^{N_\textrm{at}} \bigg\{
\sum_{j=1}^{nn(i)} \frac{3\alpha_{ij}}{4(d^0_{ij})^2}
\big[ (\mathbf{R}_j-\mathbf{R}_i)^2-(d^0_{ij})^2\big]^2 
\nn&
+\sum_{j=1}^{nn(i)}\sum_{k<j}^{nn(i)} \frac{3\beta_{ijk}}{4d_{ij}^0 d_{ik}^0}
\big[(\mathbf{R}_j-\mathbf{R}_i)\cdot(\mathbf{R}_k-\mathbf{R}_i)
\nn&
-\cos \theta_{ijk} d^0_{ij}d^0_{ik}\big]^2\bigg\},
\end{align}
where $N_\textrm{at}$ is the number of atoms, $nn(i)$ indicates the 
nearest neighbors of atom $i$, $\mathbf{R}_i$ is the position of atom $i$,
$d^0_{ij}$ is the equilibrium bond length and $\theta_{ijk}$ is the 
equilibrium bond angle between the bonds $ij$ and $jk$. Finally, $\alpha_{ij}$ 
and $\beta_{ijk}$ are the Keating parameters determining the strengths
of the bond stretching and bond bending terms, respectively.
The energy minimization is performed numerically using the conjugate-gradient
method.

\subsection{Tight-binding calculation}
The single-particle states are calculated within a tight-binding approach.
Here, we use the spds$^*$ model with $N_\textrm{orb}=20$ 
local orbitals per atom and with nearest-neighbor hopping. 
The corresponding tight-binding Hamiltonian is
\begin{align}
H_{TB}&=\sum_{i=1}^{N_\textrm{at}}\sum_{\alpha=1}^{N_\textrm{orb}}
\epsilon_{i,\alpha} c^\dagger_{i,\alpha}c_{i,\alpha}
+\sum_{i=1}^{N_\textrm{at}}\sum_{\alpha,\beta=1} ^{N_\textrm{orb}}
\lambda_{i,\alpha,\beta} c^\dagger_{i,\alpha}c_{i,\beta}
\nn&
+\sum_{i=1}^{N_\textrm{at}} \sum_{j=1}^{nn(i)} \sum_{\alpha,\beta=1}^{N_\textrm{orb}}
t_{i,\alpha,j,\beta}c^\dagger_{i,\alpha} c_{j,\beta},
\end{align}
where $\epsilon_{i,\alpha}$ is the onsite energy at orbital $\alpha$ on atom
$i$, $t_{i,\alpha,j,\beta}$ is the hopping matrix element between orbitals
$\alpha$ and $\beta$ on atoms $i$ and $j$, respectively, and 
$\lambda_{i,\alpha,\beta}$ describes the spin-orbit interaction
\cite{Chadi}. 
Using the Slater-Koster rules\cite{SlaterKoster}, the hopping elements 
$t_{i,\alpha,j,\beta}$ are calculated from a reduced 
number of hopping elements, e.~g. $V_{sp\sigma}^{ac}$ for the $\sigma$-bond between
the $s$ orbital of an anion and the $p$ orbital of a cation, combined with 
the cosines of the bond angles. Furthermore, for strained structures, we
use a generalized Harrison's law and scale the hopping elements 
$t_{i,\alpha,j,\beta}$ by a factor $(d_0/d)^{\eta_{\alpha,\beta}}$, where
$d_0$ is the equilibrium bond length, $d$ is the bond length in the strained
structure, and $\eta_{\alpha,\beta}$ is the exponent. Similarly, the diagonal energies have to be corrected in the
presence of strain, where we use the approach introduced in 
Ref.~\onlinecite{DiagonalStrainCorrection}
\begin{align}
\epsilon_{i,\alpha}=\epsilon^{eq}_{i,\alpha}+\sum_{j=1}^{nn(i)}\sum_\beta
C_{i\alpha,j\beta}\frac{(t_{i\alpha,j\beta}^{eq})^2-(t_{i\alpha,j\beta})^2}
{(\epsilon_{i\alpha}-\epsilon_\textrm{ref})+
(\epsilon_{j\beta}-\epsilon_\textrm{ref})},
\end{align}
where $\epsilon^{eq}_{i,\alpha}$ and $t_{i\alpha,j\beta}^{eq}$ refer to
the parameters for an unstrained material, $\epsilon_\textrm{ref}$ is a reference
energy value which we take to be two Rydbergs, $\epsilon_\textrm{ref}=27$ eV,
and $C_{i\alpha,j\beta}$ are the parameters determining the strengths of the
diagonal corrections.

The diagonalization of the tight-binding Hamiltonian yields the 
single-particle energies as well as the eigenvectors in the form of the
coefficients $F(i,k,\alpha)$ of an expansion of the single-electron
wave function $\Phi_i(\mathbf{r})$ in terms of the local orbitals 
\begin{align}
\Phi_i(\mathbf{r})=
\sum_{k=1}^{N_\textrm{at}}\sum_{\alpha=1}^{N_\textrm{orb}}
F(i,k,\alpha)\phi_\alpha(\mathbf{r}-\mathbf{R}_k).
\label{spstate}
\end{align}
Here, $\phi_\alpha(\mathbf{r}-\mathbf{R}_k)$ is the local atomistic
orbital centered at atom $k$. 
When needed, we approximate the radial part of these orbitals 
by the Slater formula, however its explicit functional form is not
essential for the construction of the single-particle Hamiltonian.

\subsection{Many-body calculation}
The optical properties of quantum dots are determined by excitons,
trions, and other excitonic complexes. To calculate these complexes 
one has to diagonalize the many-body Hamiltonian
\begin{align}
H_{MB}&=\sum_i E^{(e)}_i c^\dagger_i c_i 
+\frac 12\sum_{ijkl}\langle ij|V_{ee}|kl\rangle c^\dagger_ic^\dagger_jc_kc_l
\nn&
+ \sum_p E_p^{(h)} h^\dagger_p h_p
+\frac 12\sum_{pqrs}\langle pq|V_{hh}|rs\rangle h^\dagger_ph^\dagger_qh_rh_s
\nn&
-\sum_{iqrl}\big( \langle iq|V_{eh}^{\textrm{dir}}|rl\rangle 
- \langle iq|V_{eh}^{\textrm{exc}}|lr\rangle \big)
c^\dagger_ih^\dagger_qh_rc_l.
\label{h_mb}
\end{align}
Here, $c^\dagger_i (c_i)$ and $h^\dagger_p (h_p)$ refer to creation 
(annihilation) operators of electrons and holes, respectively, in the basis
of the single-particle eigenstates of the tight-binding Hamiltonian.
$E^{(e)}_i$ is the energy of the $i$-th electron state and
$E^{(h)}_p$ is the energy of the $p$-th hole state, i.e. the negative of
the eigenvalue of the tight-binding Hamiltonian for states below the gap.
Furthermore, the many-body Hamiltonian contains the electron-electron 
($V_{ee}$), hole-hole ($V_{hh}$), and electron-hole direct 
($V_{eh}^\textrm{dir}$) and exchange ($V_{eh}^\textrm{exc}$) 
terms of the Coulomb interaction.

To enable the calculation of Coulomb matrix elements for system with 
about one million atoms, some approximations have to be made. In particular,
we calculate only two-center terms and treat onsite and long-range terms 
differently. The onsite Coulomb matrix element between Slater orbitals 
can be precomputed for each material. 
For the long-range terms we can replace
the position difference $|\mathbf{r}_1-\mathbf{r}_2|$ of electrons or holes 
by the difference 
of the positions of the atoms around which orbitals are centered and apply
orthogonality relations to solve the remaining integrals.
For example, for the electron-electron term, we arrive at
the onsite (OS) and long-range (LR) contributions
\begin{subequations}
\begin{align}
&\langle ij| V^{(OS)}_{ee} |kl\rangle=
\frac{e^2}{4\pi\epsilon_0 \epsilon_{OS}} 
\nn& \times
\sum_{a=1}^{N_\textrm{at}}
\sum_{\alpha,\beta,\gamma,\delta=1}^{N_\textrm{orb}}
F^*(i,a,\alpha)F^*(j,a,\beta)F(k,a,\gamma)F(l,a,\delta)
\nn& \times
\bigg[\int d^3r_1 \int d^3r_2
\frac{\phi^*_\alpha(\mathbf{r}_1) \phi^*_\beta(\mathbf{r}_2)
\phi_\gamma(\mathbf{r}_2)\phi_\delta(\mathbf{r}_1)}
{|\mathbf{r}_1-\mathbf{r}_2|}\bigg], \\
&\langle ij| V^{(LR)}_{ee} |kl\rangle=
\frac{e^2}{4\pi\epsilon_0 \epsilon_{LR}}
\sum_{a_1=1}^{N_\textrm{at}} 
\sum_{\substack{a_2=1 \\ a_2\neq a_1}}^{N_\textrm{at}}
\sum_{\alpha,\beta=1}^{N_\textrm{orb}}
\frac{1}{|\mathbf{R}_{a_1} - \mathbf{R}_{a_2}|}
\nn& \times
F^*(i,a_1,\alpha)F^*(j,a_2,\beta)F(k,a_2,\beta)F(l,a_1,\alpha).
\end{align}
\end{subequations}
Here, we assume the long-range terms are screened by the bulk dielectric 
constant $\epsilon_{LR}=\epsilon$ while the onsite terms are taken as
unscreened $\epsilon_{OS}=1$.

With the Hamiltonian (\ref{h_mb}) fully parametrized, we can now compute
the ground and excited states of four fundamental excitonic complexes:
the neutral exciton X, the positively and negatively charged exciton
X$^+$ and X$^-$, respectively, and the biexciton XX.
In each case we form the many-body basis by generating all possible
configurations of the electrons and holes on a chosen set of 
single-particle states.
Specifically, the configurations for $X$ take the form $|ip\rangle = 
c_i^+h_p^+|0\rangle$, where $|0\rangle$ denotes the vacuum state
(an empty quantum dot).
For $X^+$ these configurations have the form 
$|ipq\rangle = c_i^+h_p^+h_q^+|0\rangle$, while for $X^-$ they are
$|ijp\rangle = c_i^+c_j^+h_p^+|0\rangle$.
Finally, the biexciton states are built in the basis
$|ijpq\rangle = c_i^+c_j^+h_p^+h_q^+|0\rangle$.
In all these basis configurations, the indices $i,j$ ($p,q$) enumerate 
all available electron (hole) single-particle states obeying
Fermionic occupation rules.

For each excitonic complex, diagonalization of the Hamiltonian
(\ref{h_mb}) in the appropriate basis gives us the ground and
excited states in the form of linear combinations
\begin{align}
    |X_{\alpha}\rangle = \sum_{i,p} A_{i,p}^{\alpha} |ip\rangle
\end{align}
for the exciton, and analogous forms for the other complexes.
Here the index $\alpha$ enumerates the excitonic states 
and $A_{i,p}^{\alpha}$ are the expansion coefficients of the eigenvector.

\subsection{Optical spectra}

One of the ingredients in the calculation of the emission spectra
of the excitonic complexes is the dipole element
\begin{align}
    D_{ip}(\boldsymbol{\epsilon}) 
    = \langle i | \mathbf{r}\cdot\boldsymbol{\epsilon} | p \rangle,
\end{align}
where $\mathbf{r}$ is the position operator, while
$\boldsymbol{\epsilon}$ denotes the polarization of emitted light.
This element is computed with the single-particle states of the form as
in Eq. (\ref{spstate}), respectively for the electron ($i$)
and the hole ($p$).
Details of calculation of this element in the tight-binding basis
are given in Ref.~\onlinecite{Marek_review}.
Utilizing the elements $D_{ip}$ we now define the interband
polarization operator
\begin{align}
    P(\boldsymbol{\epsilon}) = \sum_{ip}D_{ip}c_ih_p
\end{align}
which removes one electron-hole pair from the system
obeying optical selection rules.
This operator is central in computing the emission spectra
from the state $|\alpha\rangle$ of one of our four excitonic complexes
(with $N_e$ electrons and $N_h$ holes).
The final state $|f\rangle$ in such a transition is the correlated
ground or excited state of the system with $N_e - 1$ electrons and 
$N_h - 1$ holes.
The emission spectrum $F(E,\boldsymbol{\epsilon})$ as a function of
the energy $E$ and light polarization $\boldsymbol{\epsilon}$
is computed using the Fermi's golden rule
\begin{align}
F(E,\boldsymbol{\epsilon}) = F_0
\sum_{\alpha,f} | \langle \alpha | P(\boldsymbol{\epsilon}) | f \rangle|^2
\delta\big[E-(E_f-E_{\alpha})\big] n_{\alpha} (1-n_f),
\end{align}
where $F_0$ is a constant depending on the light-matter interaction,
$E_{\alpha}$ and $E_f$ are the energies of the manybody states 
involved in the transition, and
$n_{\alpha}$ and $n_f$ denote the occupations of the respective states.
In what follows we focus on the total (unpolarized) emission calculated
as the sum $F(E):=F(E,x)+F(E,y)+F(E,z)$ of all possible polarizations.
For simplicity we assume that the final states are unoccupied 
($n_f=0$) and that due to thermalization 
only the $i_0$ lowest-energetic initial states are occupied 
($n_{\alpha}=1$ for $\alpha\le i_0$ and $n_{\alpha}=0$ for $\alpha>i_0$). 
Throughout this article, we focus on the emission predominantly originating 
from the lowest s-type states, i.e., $i_0=4$ for excitons, $i_0=1$ for 
biexcitons, and $i_0=2$ for trions. 
Note that, because the trions consist of an odd number of
fermions, the Kramers theorem applies and also the manybody states of trions
come in degenerate pairs. Here, we choose to work with those linear 
combinations of trion states that minimize (maximize) the spin of the 
unpaired charge carrier.

\subsection{Numerical implementation}
In order to be able to use the computational tools developed here not only for 
the present study of single quantum dots but more generally quantum dot arrays 
in nanowires, which eventually requires the calculation of systems with millions
of atoms, we have completely refactored the toolkit QNANO~\cite{Marek_review,
Zielinski_selfassembled_InAs,Weidong_selfassembled_InAs,Weidong_LR}. 
We have parallelized the central parts using MPI so that calculations can be 
performed on hundreds of cores on high-performance computers.
For the diagonalization of matrices such as the tight-binding and the 
manybody Hamiltonian, we use the highly parallelized Krylov-Schur algorithm 
implemented in the PETSc\cite{PETSC1,PETSC2}-based library SLEPc\cite{SLEPC}.

One of the most time consuming steps in our procedure is the calculation 
of the Coulomb matrix elements. The onsite terms can be distributed 
straightforwardly as the contributions from different atoms are independent.
The long-range terms can be cast into a form that allows for an efficient
calculation in terms of matrix-vector multiplications as layed out in 
Ref.~\onlinecite{Weidong_LR}. Here, we parallelize the procedure by splitting
the respective matrix into blocks that are calculated and stored on different
nodes.

\section{Band structure calculations}
In order to develop a predictive tool based on the empirical tight-binding 
method, one would like to obtain the hopping elements, diagonal energies, 
and strain correction parameters from empirical data.
However, in contrast to bulk InAs and InP samples, which grow in the 
zincblende crystal phase, here, we want to model nanowires in the wurtzite 
phase, where experimental data, e.g., on the band
structure of InAs and InP, is scarse. 

Although locally zincblende and wurtzite phases are very similar in that
they are tetragonally coordinated with first differences appearing in 
third-nearest neighbors, significant differences in the band structures
are expected: The wurtzite phase shows a crystal field splitting that 
is absent in zincblende structures and the fundamental energy gap is larger
in wurtzite structure by, e.g., $59$ meV for InAs\cite{wzInAs_Rota}.
These differences preclude a direct reuse of established tight-binding 
parameters for zincblende InAs and InP \cite{Jancu1998}.

In order to obtain a set of tight-binding parameters together with the
strain correction parameters for wurtzite InAs and InP, 
we use a fitting precedure to reproduce band structures in bulk systems.
The band structures are obtained by \emph{ab initio} methods. 
We perform DFT calculations using the PBE energy functional in a plane wave
basis with full relativistic PAW pseudopotentials\cite{DalCorso2012}
within the QUANTUM ESPRESSO \cite{QUANTUM_ESPRESSO1, QUANTUM_ESPRESSO2} code.
The calculations are performed for several different strained unit cell 
configurations so that the strain correction parameters can be obtained in 
a simultaneous fit of multiple band structures.

As DFT calculations underestimate the band gap significantly, we 
perform a scissors shift of the conduction band states to the empirically
known fundamental gaps of $E_g = 1.490$ eV in wurtzite InP
\cite{wzInP_Dalacu_2014} and $E_g = 0.477$ eV in wurtzite InAs 
\cite{wzInAs_Rota}, respectively.
In fact, because the gap in InAs is small, the underestimation of the gap 
in calculations using the PBE functional leads to the closing of the gap.
Therefore, we focus the fitting to compressed wurtzite InAs structures 
with lattice constants close to that of the InP matrix, which is about $3\%$
smaller. Note that the dot region in nanowire quantum dot systems 
is typically composed of InAs$_x$P$_{1-x}$ with $x\sim 20\%$. Due to the
relatively low As concentration, the As atoms will on average experience an
environment with lattice constants closer to that of InP than of InAs. 
Thus, the compressed InAs lattice is a reasonable starting point for the fitting
when the parameters will be used to model nanowire quantum dots.

Furthermore, the zero of energy is ill-defined in calculations with 
periodic boundary conditions containing Coulomb interaction terms 
proportional to $\frac 1r$. Thus, the band alignments between 
InAs and InP as well as the alignments for different strain configurations 
are much more difficult to calculate from first principles. 
Here, we take the zincblende values of
the absolute deformation potentials for the valence band maxima of 
$a_V^{VBM} =1.83$ eV for InP and 
$a_V^{VBM} =1.79$ eV for InAs from Ref.~\onlinecite{AbsDefPot}
and the natural band offset between InAs and InP at their respective 
equilibrium lattice constants of
$\Delta E_v^{eq}(\textrm{InAs})-\Delta E_v^{eq}(\textrm{InP})=0.47 $ eV
from Ref. \onlinecite{NaturalBandOffsets}.

\begin{figure}
\includegraphics[width=\linewidth]{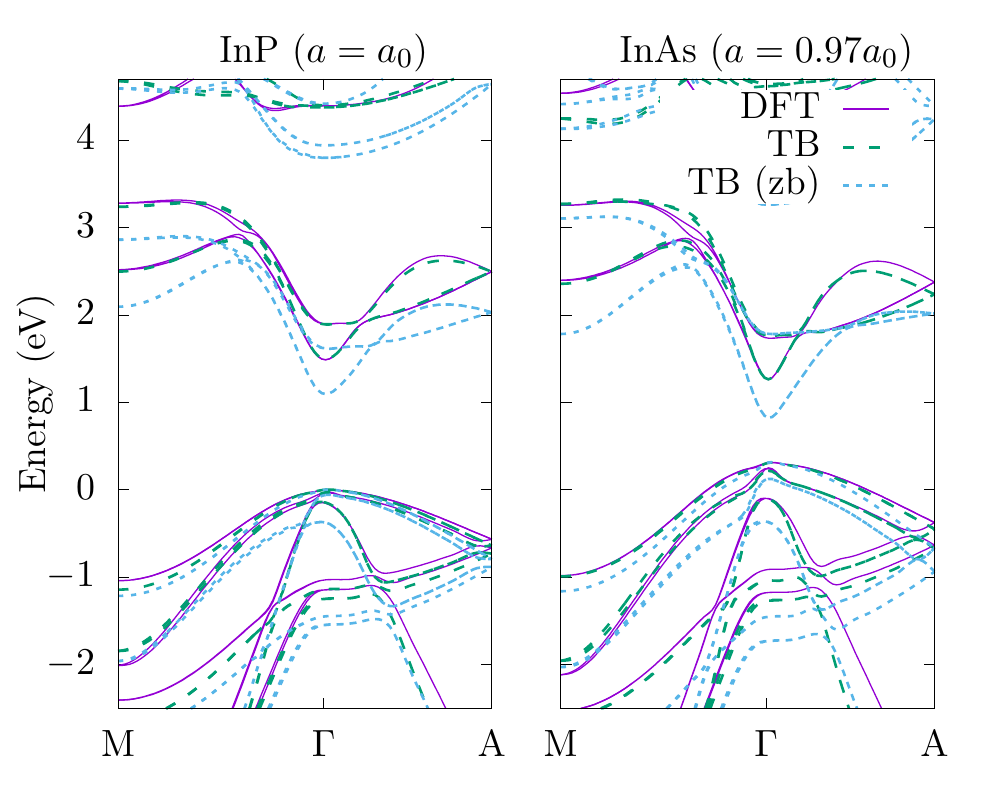}
\caption{\label{fig:plt_fit} Band structure of InP (left) in the wurtzite phase
at its equilibrium lattice constant $a=a_0$ and of InAs (right) 
in a hydrostatically compressed wurtzite crystal with lattice constant 
$a=0.97 a_0$. 
The DFT calculations have been corrected by scissors shifts. The 
tight-binding (TB) calculation are obtained from the fitted parameters.
For comparison we show tight-binding results [TB (zb)] obtained 
from the zincblende parameters from Ref.~\onlinecite{Jancu1998}, where
the energies are shifted so that the valence band maxima coincide with that
of the DFT calculations. 
}
\end{figure}

Note that there is a significant variation in the literature values of
the absolute deformation potentials\cite{AbsDefPot_Kadantsev}. Also,
DFT calculations reveal that $a^{\textrm{gap}}_v=dE^\textrm{gap}/d\ln(V)$ deviates
visibly from a constant over the range of changes of $3\%$ of the lattice 
constants. For example, including the scissors shift we obtain the energy
gap of wurtzite InAs at $3\%$ compression as $E_g^{-3\%}=0.960$ eV, 
while starting from the corrected gap at equilibrium and adding the 
contribution of the relative deformation potential of the gap 
$a_v^{CMB}-a_v^{VMB}$ from Ref.~\onlinecite{AbsDefPot} yields a value of
$E_g^{-3\%}=0.991$ eV. Thus, uncertainty in deformation potentials and 
band alignments defines the order of magnitude of the error of absolute 
energies in our strained band structure calculations
and can be estimated as $\sim 30$ meV.

\section{Parameter fitting}
\begin{table}[h!]
\begin{tabular}{c@{\hskip 3mm}c}
\begin{tabular}[t]{l|c|c}
& InP (eV) & InAs (eV)\\
\toprule
$E_s^{a}$ & -3.9798  & -5.3673 \\
$E_s^{c}$ & -1.9268  & -0.9905 \\
$E_p^{a}$ & 2.3067  & 2.3917 \\
$E_p^{c}$ & 6.8862  & 6.6883 \\
$E_{pz}^{a}$ & 2.3389  & 2.6082 \\
$E_{pz}^{c}$ & 6.2922  & 6.0555 \\
$E_{d}^{a}$ & 13.4134  & 13.7658 \\
$E_{d}^{c}$ &  12.2801 & 12.5896 \\
$E_{s^*}^{a}$ & 19.2302  & 19.5859 \\
$E_{s^*}^{c}$ & 19.1728 &  18.3726 \\
$\Delta_{so}^a$ & 0.0217  & 0.1482 \\
$\Delta_{so}^c$ & 0.1675  & 0.0645 \\
\end{tabular}
&
\begin{tabular}[t]{l|c|c}
& InP (eV) & InAs (eV)\\
\toprule
$V_{ss\sigma}$ & -2.6537 & -3.5352 \\
$V_{sp\sigma}^{ac}$ & 3.3428 & 4.3074 \\ 
$V_{sp\sigma}^{ca}$ & 3.3557 & 3.5224 \\
$V_{pp\sigma}$ & 3.8437 & 3.5053 \\
$V_{pp\pi}$  & -1.2305 & -1.2792 \\
$V_{ss^*\sigma}^{ac}$ & -1.3517 & -2.5597\\
$V_{ss^*\sigma}^{ca}$ & -3.8673 & -5.5423 \\
$V_{s^*s^*\sigma}$ & -4.3416 & -5.7421 \\
$V_{s^*p\sigma}^{ac}$ & 2.6263 & 2.2262 \\
$V_{s^*p\sigma}^{ca}$ & 3.1080 & 3.5650 \\
$V_{sd\sigma}^{ac}$ & -3.2625 & -4.7604 \\
$V_{sd\sigma}^{ca}$ & -2.4241 & -2.9015 \\
$V_{pd\sigma}^{ac}$ & -1.8250 & -1.6637\\
$V_{pd\sigma}^{ca}$ & -1.3002 & -1.2407 \\
$V_{pd\pi}^{ac}$ & 1.4239 &  1.3719 \\
$V_{pd\pi}^{ca}$ & 1.6504 & 2.2598 \\
$V_{s^* d\sigma}^{ac}$ & -0.7779 & -1.2618 \\
$V_{s^* d\sigma}^{ca}$ & -0.6759 & -1.1917 \\ 
$V_{dd\sigma}$ & -1.8423 & -2.3460 \\
$V_{dd\pi}$ & 3.2696 & 2.4663\\
$V_{dd\delta}$ & -0.5511 & -0.9812
\end{tabular}
\end{tabular}

\vspace{0.5cm}

\caption{\label{tab:params}Tight-binding parameters for InP and InAs in 
the wurtzite phase.}
\end{table} \begin{table}[h]
\begin{tabular}{c@{\hskip 3mm}c}
\begin{tabular}[t]{l|c|c}
& InP & InAs \\
\toprule
$\eta_{ss\sigma}$ & 2.5931 & 1.7590 \\
$\eta_{sp\sigma}^{ac}$ & 1.6599 & 2.0673 \\
$\eta_{sp\sigma}^{ca}$ & 2.0125 & 2.9042 \\
$\eta_{pp\sigma}$ & 3.4803 & 4.4087 \\
$\eta_{pp\pi}$  & 2.1529 & 1.7411 \\
$\eta_{ss^*\sigma}^{ac}$ & 2.3711 & 2.7047\\
$\eta_{ss^*\sigma}^{ca}$ & 1.5276 & 2.2829 \\
$\eta_{s^*s^*\sigma}$ & 2.0682 & 1.8149 \\
$\eta_{s^*p\sigma}^{ac}$ & 2.4391 & 1.7442 \\
$\eta_{s^*p\sigma}^{ca}$ & 2.7750 & 2.5430 \\
$\eta_{sd\sigma}^{ac}$ & 2.6314 & 1.7371 \\
$\eta_{sd\sigma}^{ca}$ & 2.0122 & 2.1701\\
$\eta_{pd\sigma}^{ac}$ & 1.7862 & 1.6910 \\
$\eta_{pd\sigma}^{ca}$ & 1.5977 & 1.9338 \\
$\eta_{pd\pi}^{ac}$ & 1.8807 & 1.7122 \\
$\eta_{pd\pi}^{ca}$ & 2.2384 & 2.5487\\
$\eta_{s^* d\sigma}^{ac}$ & 1.8903 & 1.4862 \\
$\eta_{s^* d\sigma}^{ca}$ & 2.1164 & 2.0633 \\
$\eta_{dd\sigma}$ & 2.3844 & 2.5827 \\
$\eta_{dd\pi}$ & 2.3570 & 3.1018 \\
$\eta_{dd\delta}$ & 2.3391 & 2.7207
\end{tabular}
&
\begin{tabular}[t]{l|c|c}
& InP & InAs \\
\toprule
$C_{ss}$ &1.7460 &2.8083\\
$C_{sp}^{ac}$ & 3.8895 & 4.6150 \\
$C_{sp}^{ca}$ & 4.3656 & 4.4938 \\
$C_{pp}$ & 0.9416 & 0.8099 \\
$C_{ss^*}^{ac}$ & 0.2023 & -0.7137 \\
$C_{ss^*}^{ca}$& -0.4574 & -0.7256 \\
$C_{ps^*}^{ac}$ & -0.4921 & 0.4052 \\
$C_{ps^*}^{ca}$ & 0.1400 & -0.5372 \\
$C_{s^*s^*}$ & -0.4994 & -1.3230 \\
$C_{sd}^{ac}$ & 0.1546 & 0.7472 \\
$C_{sd}^{ca}$ & -0.3382 & 0.6127 \\
$C_{pd}^{ac}$ & -0.5661 &-1.2332 \\
$C_{pd}^{ca}$ & -0.0966 & 0.3289 \\
$C_{s^*d}^{ac}$ & -0.4380 & -0.9437 \\
$C_{s^*d}^{ca}$ & 0.6783 &  0.2654 \\
$C_{dd}$ & -0.8597 &-1.5460
\end{tabular}
\end{tabular}
\caption{\label{tab:strain_params}Strain parameters: 
Harrison's law exponents $\eta$ and diagonal correction coefficients $C$.}
\end{table}

The fitting of the band structures from the M point to $\Gamma$ to the A point
is performed using a conjugate gradient method with stochastic
basin hopping to avoid trapping in local minima.
For each material, there are 33 tight-binding parameters
for structures in absence of stain and 37 additional 
strain correction parameters. Due to this large parameter space, many 
different parametrizations can yield similarly good fits. Also, the target 
function to be optimized is not unambiguous. For example, one can select
different bands for the fit or attribute different weights to different bands.
The physics in direct band gap materials is largely determined 
by the k-space region close to the $\Gamma$ point, so the fitting procedure
should weight values close to $\Gamma$ more strongly than other points in
the Brillouin zone. We also fix the signs of the most important hopping 
elements, penalize strong deviations of the generalized Harrison's law
coefficients $\eta_{i,\alpha}$ from 2 and suppress large values of the 
diagonal strain corrections. Note also that fitting the band structure 
through multiple points in k-space together with defined symmetries of the
local orbitals enables a distinction between terms originating from s, p or
d orbitals. However, the s$^*$ orbital, which is designed to model higher 
lying conduction band states, has the same symmetry as the s orbital, so that
s and s$^*$ orbitals cannot be distinguished by symmetry in a naive fitting 
procedure of the band structure. 
We therefore design the target function to penalize small values of the 
s$^*$ diagonal energy. Finally, to ensure compatibility between the 
tight-binding parameters for wurtzite InAs and InP, we first obtain the InP
parameters and then we use those values as a starting guess for the fit for 
InAs while reducing the range of the stochastic basin hops to remain
in a minimum compatible to the InP parameters.

Figure \ref{fig:plt_fit} shows the band structures in the wurtzite phase
for InP at its equilibrium
lattice constant and InAs with a lattice constant compressed by 3\% of its
equilibrium value 
from the M point through $\Gamma$ to the A point, where scissor shifts 
have been applied to correct the DFT gap to reproduce the experimental values.
The results of the tight-binding calculations for the fitted tight-binding 
parameters as well as calculations with zincblende InP parameters from 
Ref.~\onlinecite{Jancu1998} are compared, where in the latter calculation 
all energies have been shifted so that the valence band maximum coincides 
with that of the DFT calculations. The fit reproduces the overall 
DFT band structure well. The calculation using the zincblende parameters 
on the other hand predicts a smaller band gap and shows qualtitative different
behaviour especially around the valence band maximum due to the lack of a
crystal field splitting. 

The tight-binding parameters for InP and InAs
in the wurtzite phase obtained from our fitting procedure 
are listed in table \ref{tab:params}. 
The strain parameters, i.e. the Harrison's law exponents $\eta$ 
and the diagonal strain corrections $C$, are shown in table 
\ref{tab:strain_params}.
The labels $a$ and $c$ refer to the anion (P/As) and the cation (In), 
respectively. Except for the spin-orbit splitting parameters 
$\Delta^a_{so}$ and $\Delta^c_{so}$, which are relevant for the p-orbitals,
the indices refer to the local orbital or the bonds.
For example, the onsite energy of the s-orbital on an In atom is
$E^c_s$, the Slater-Koster hopping element\cite{SlaterKoster} between
between an s-orbital on P and a p-orbital on an In atom in InP forming a 
$\sigma$-bond is denoted by $V^{ac}_{sp\sigma}$. Because of the hermiticity of the
tight-binding Hamiltonian, $V^{ac}_{sp\sigma}=V^{ca}_{ps\sigma}$, and 
$V^{ca}_{ps\sigma}$ is not listed explicitly.

\section{Results}
Having obtained the tight-binding parameters, we now move on to calculate
single-particle states and spectra of typical hexagonal InAs$_x$P$_{1-x}$
nanowire quantum dots and investigate the influence of random alloying 
and changes of sizes and As concentrations.

\subsection{Single-particle states and spectral lines for typical InAsP
nanowire quantum dots}
\begin{figure}
\includegraphics[width=0.95\linewidth]{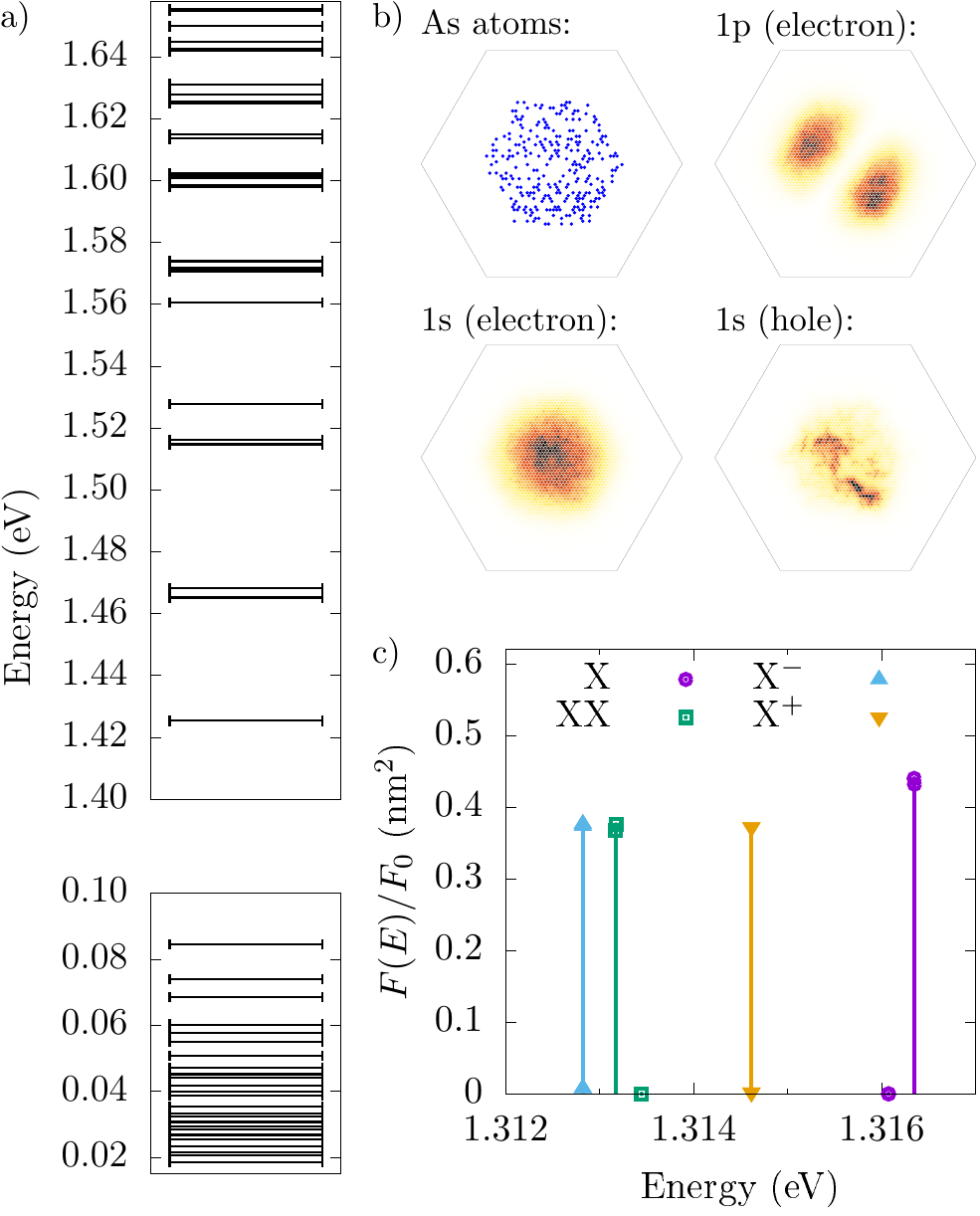}
\caption{\label{fig:SP} Single-particle energy eigenstates for a hexagonal
InAs$_{0.2}$P$_{0.8}$ quantum dot with diameter $D\approx 18$ nm and height $d\approx 4$ nm
in an InP environment. The energy levels closest to the gap are depicted in 
a). b) shows a horizontal cut though the quantum dot, the distribution of
As atoms as well as the densities of single-particle states close
to the gap. The spectral lines originating from the lowest-energetic
excitons, biexcitions and trions are depicted in c).
}
\end{figure}

A typical hexagonal InAs$_x$P$_{1-x}$ nanowire quantum dot as grown 
and described in Ref.~\onlinecite{Dalacu_pureWZ} has a diameter of 
$D\approx 18$ nm, a height of $h\approx 4$ nm, and an As concentration 
of $x=20\%$. 
We first present results on one realization of such a dot in Fig.~\ref{fig:SP}.
The distribution of As atoms in a horizontal cross section through the 
computational box is depicted in the top left panel of Fig.~\ref{fig:SP}b.
For all calculations presented here, we use a computational box size containing
about 400,000 atoms. For the manybody problem we compute long-range Coulomb 
matrix elements between 40 electron and hole states and add onsite Coulomb
terms between 12 single-particle states per band. Convergence is discussed
in detail in section \ref{sec:convergence}.

Figure \ref{fig:SP}a shows the single-particle energy eigenvalues around the
gap for this quantum dot, where the energy is defined with respect to the 
equilibrium bulk wurtzite InP valence band maximum. 
Note that each line corresponds to two Kramers degenerate pairs due to 
time-reversal symmetry.
A clear shell structure is visible around the gap.
For a two-dimensional dot with parabolic confinement one expects s-type states
followed by degenerate p-type states.
To verify that the same character of the states is also obtained in the 
atomistic calculation, we show in Fig.~\ref{fig:SP}b 
the density corresponding to the lowest energetic conduction band state at
energy 1.425 eV (1s electron), the conduction band state at 1.465 eV 
(1p electron) and the highest valence band state at energy 0.084 eV (1s hole).
While
the wave functions in the conduction band are relatively smooth and symmetric,
the valence band states show some granularity. This is due to the fact that
the natural band alignment between InP and InAs leads to a shallow confining 
potential for conduction band electrons whereas the holes experience 
a deep confining potential, 
so that the random incorporation of As atoms has a much more drastic
impact on the wave functions for holes.

In Fig. \ref{fig:SP}c, the predicted spectral lines originating from the
lowest four exciton states (X), the lowest biexciton state (XX) and the
lowest two (Kramers degenerate) trion states (negative X$^-$ and positive X$^+$)
are presented. 
For this particular quantum dot, the exciton (X) spectra depicted 
in Fig. \ref{fig:SP}c reveal two dark states 
about 0.3~meV below two bright lines. The splitting between the two
dark exciton states is about 0.2~$\mu$eV while the fine structure splitting 
(FSS) of the bright excitons is about 3~$\mu$eV.

The lines originating from the biexciton (XX) mirror the behavior 
of the exciton lines. The biexciton binding energy 
$\Delta E_B=2 E_{\bar{X}}-E_{XX}$ with respect to 
the  average energy $E_{\bar{X}}$ of the two bright exciton states 
is about $\Delta E_B=3$~meV. The trion lines are found to the left and to the
right of the biexciton line, where the emission energy of the negative trion 
is lower than that of the positive trion. This can be traced back to the 
fact that the holes are more strongly confined than electrons and, thus,
the Coulomb repulsion between holes is stronger than the repulsion between
electrons.

\subsection{Effects of random As incorporation}
\begin{figure}
\includegraphics[width=0.95\linewidth]{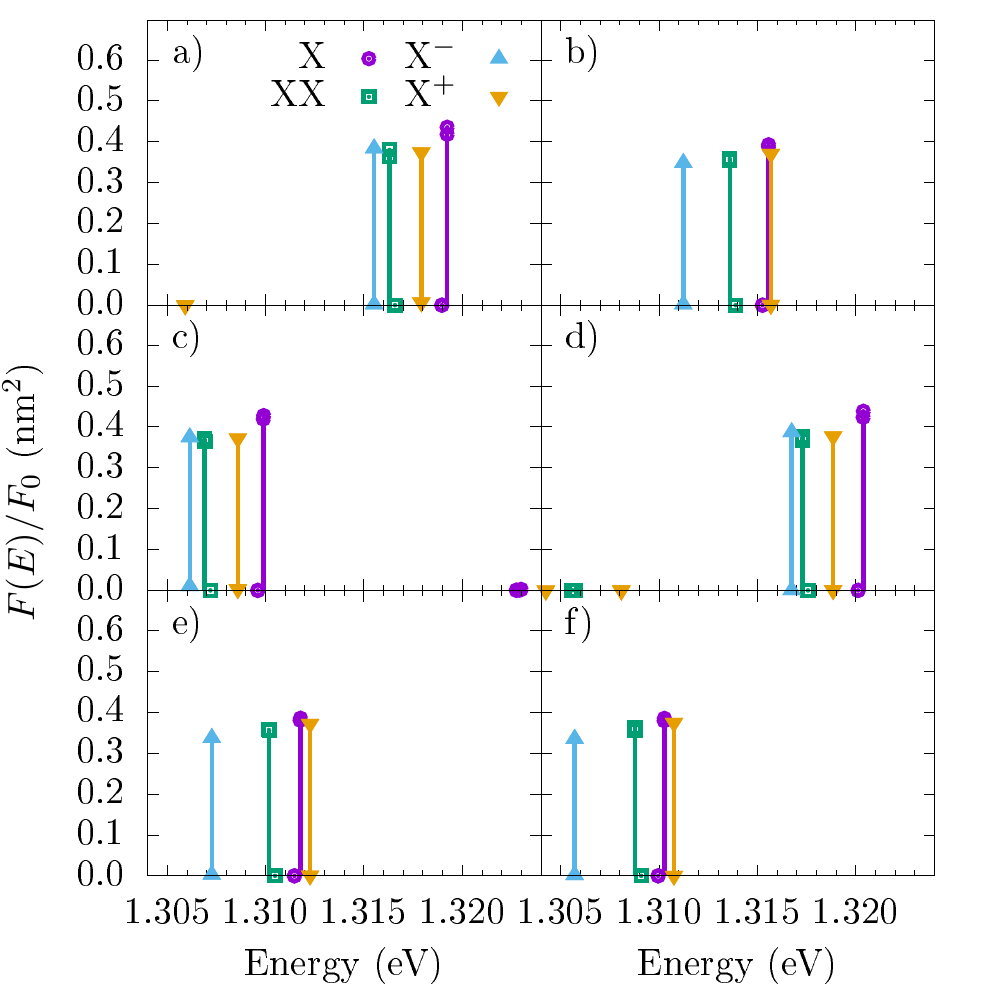}
\caption{\label{fig:random} Spectra of the lowest excitonic complexes for 
different random realizations of InAs$_{0.2}$P$_{0.8}$ quantum dots
with diameter 18 nm and height 4 nm.
}
\end{figure}

\begin{figure*}
\includegraphics{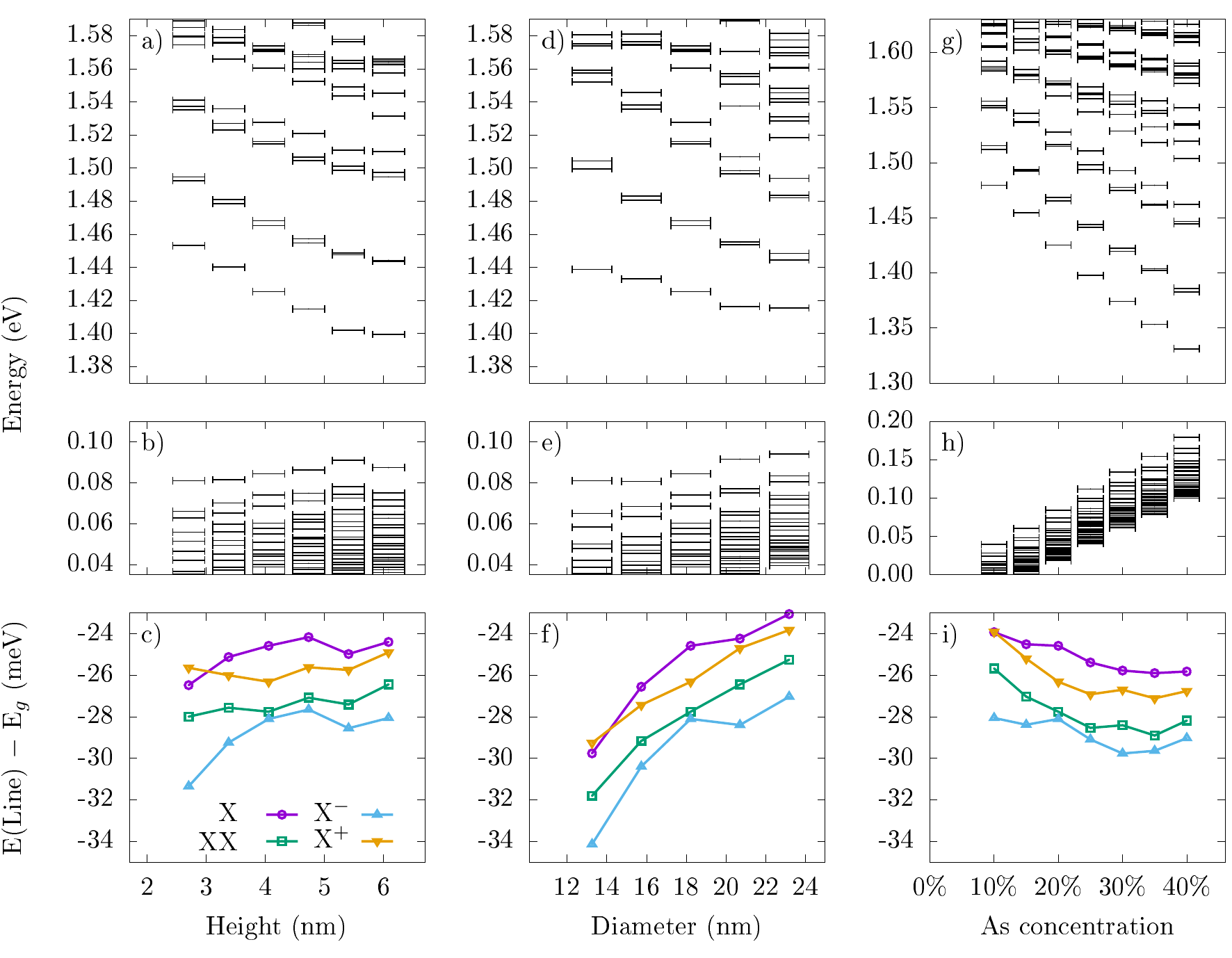}
\caption{\label{fig:geom} Dependence of single-particle conduction
(first row) and valence band states (second row) on the 
height (first column), on the diameter (second column) and on the 
As concentration within the quantum dot (third column).
The bottom row shows the positions of the exciton (X), 
biexciton (XX) and negative (X$^-$) and positive (X$^+$) trion lines 
with respect to the single-particle gap $E_g$.
}
\end{figure*}

Due to the random alloying of As within the InP matrix in an 
InAs$_x$P$_{1-x}$ nanowire quantum dot the details of the electronic
and optical properties fluctuate from one realization 
of a quantum dot to another.
To investigate the effects of randomness we show in Fig.~\ref{fig:random}
the spectra for excitons, biexcitons and trions as in Fig.~\ref{fig:SP}c
for 6  nominally identical quantum dots with As concentration $x=20 \%$,
diameter 18 nm and height 4 nm that differ only in the locations of the 
substitutionally incorporated As atoms.

From the calculations we extract an 
average energy (standard deviation) of the 1s conduction band state 
of 1.4258 eV (1.8 meV),
whereas the highest s-type valence band state is located at 0.0857 eV 
(2.0 meV).
The average splitting (standard deviation) between the lowest s- and p-type conduction band 
states is 42.2 meV (1.5 meV).
The average splitting within the lowest conduction band 
p-type orbitals is 2.5 meV (1.0 meV), indicating that the splitting in the p 
shell is dominated by the random As distribution within the quantum dot. 

The randomness of alloying affects the manybody states in that the 
position of the bright exciton line $E(X)$ 
fluctuates with a standard deviation of 4.1 meV around the mean of 1.3148 eV. 
The largest contribution to this fluctuation originates from the 
single-particle states.
The exciton binding energy $E_g-E(X)$ and the biexciton binding energy 
$\Delta E_B$ vary with standard deviations of 0.75 meV and 0.72 meV, 
around the values of 25.3 meV and 2.4 meV, respectively.
For the average fine structure splitting of the two lowest bright excitons
we find a value of 7.1 $\mu$eV with a standard deviation of 4.6 $\mu$eV.

For the realizations of quantum dots studied here we always find the 
biexciton line at lower energies compared with the exciton line and 
the negative trion line at lower energies compared with the positive trion
line. However, whether the positive trion line is found to the left or to
the right of the exciton line varies from one random realization to another.

It is noteworthy that the relative alignment of the exciton, biexcion and trion
lines qualitatively agrees with tight-binding calculations for cylindrical
InAs/InP quantum dots in the zincblende crystal phase
\cite{Zielinski_complexes,Zielinski_dashes}.
Furthermore, empirical pseudopotential calculations comparing lense-shaped 
zincblende InAs/InP quantum dots with InAs/GaAs quantum dots\cite{Lixin_He_psp}
reveal that embedding InAs dots in a InP matrix leads to more weakly bound
biexcitons and positive trions compared with dots in a GaAs environment.
This is attributed to the different band alignments between InAs and the
respective host material, which leads to more localized hole states in 
InAs/InP. As a result, the hole-hole Coulomb repulsion is increased, which,
for excitonic complexes involving more than one hole, 
counteracts the binding due to the electron-hole interaction and 
correlation energy. A clear indication that the same effect is present in 
InAs/InP dot in the wurtzite phase is the significant splitting between
the positive and negative trion lines observed throughout all quantum dots
investigated here.

\subsection{Dependence on size and As concentration}
Figure~\ref{fig:geom} summarizes calculations for quantum dots
with varying heights, diameters and As concentrations centered around a
prototypical dot of height 4 nm, diameter 18 nm and As concentration of 
$x=20\%$. 

The single-particle energy levels show typical confinement effects, where
the single-particle gap becomes larger for smaller structures, i.e. smaller
heights or diameters. The splitting between s- and p-type states
remains nearly constant when changing the height of the dot, but it 
shrinks for increasing diameter, because the s-p-splitting is mainly 
determined by the lateral confinement.
An increased concentration of As atoms leads
to effectively deeper confining potentials, reducing the energy of 
confined particles, which also reduces the effective single-particle gap.

In order to distinguish genuine size-dependent effects of the 
manybody problem from the confinement effects of the single-particle levels,
we plot in the bottom row of Fig.~\ref{fig:geom} the position of
the average bright exciton, biexciton, and trion lines 
subtracting the respective single-particle gaps $E_g$. 
The most significant impact of the geometry on the renormalization of the 
manybody energies due to the Coulomb interaction is found for varying 
the lateral confinement, where 
the excitonic complexes tend to be more strongly bound for
quantum dots with smaller diameters. This can be explained by the fact that
the Coulomb interaction mixes states derived from the s-shell with higher
lying states. The mixing is strongly influenced by the 
energetic distance to the remote states. The states that contribute most
to the renormalization of s-derived states are the p-type states. 
Because the splitting between the s- and the p-shell increases for
decreasing diameters, the mixing becomes weaker and the carriers are
more strongly forced onto the s-shell. This, in turn, increases
the Coulomb interaction and leads to stronger renormalizations 
of the manybody energies.
In contrast, the dependence of the manybody energy renormalizations on the 
quantum dot height or the As concentration is weak. It is noteworthy that
within the range of the parameters investigated here, the relative positions 
of the exciton, biexciton, and trion lines does not depend significantly 
on the geometry of the quantum dots or the As concentration.

\begin{figure}
\includegraphics[width=0.95\linewidth]{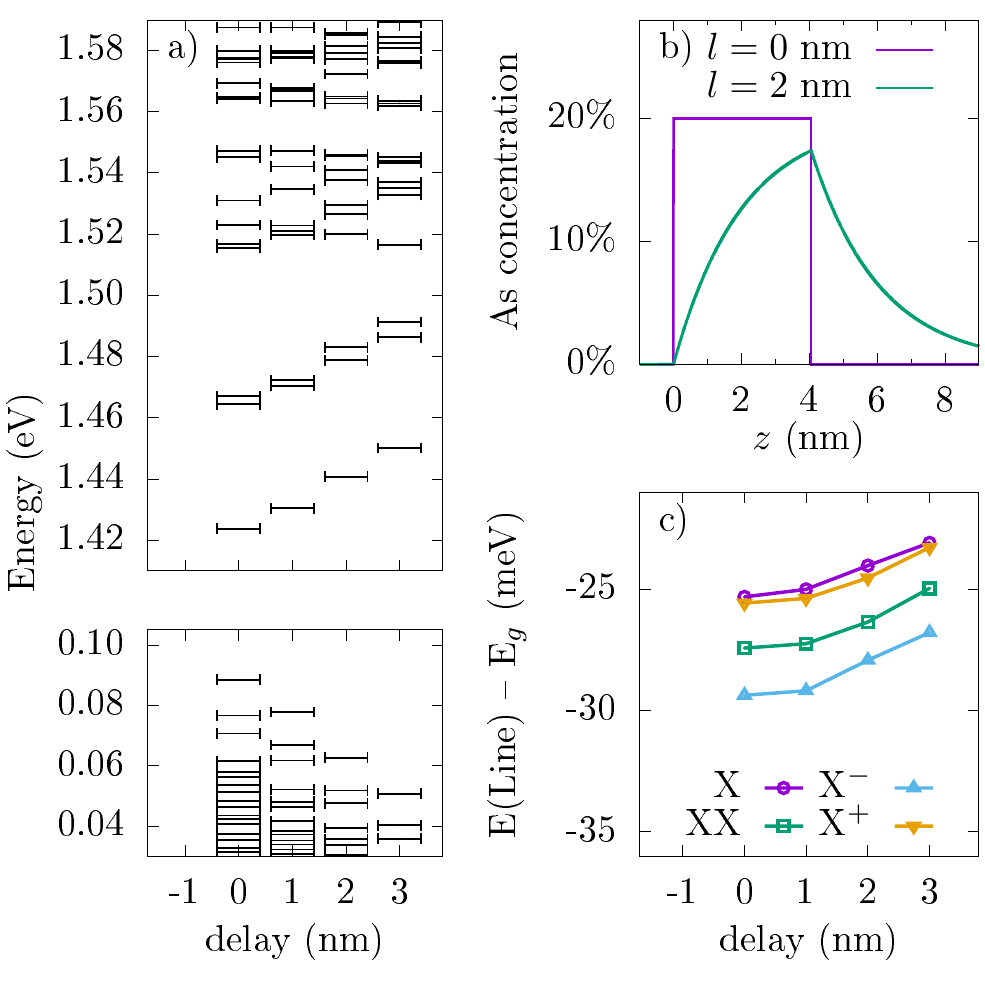}
\caption{\label{fig:delay} Delayed incorporation of As atoms (b) and
its effect on the single-particle states (a) and spectral lines
(c).
}
\end{figure}

\subsection{Delayed As incorporation}
So far, we have discussed hexagonal dots with a uniform As
distribution throughout the dot. For such a dot the As concentration 
is a step-like function of the $z$-coordinate (growth direction)
clearly distinguishing the dot region from the surrounding InP environment
(cf. concentration for $l=0$ nm in Fig.~\ref{fig:delay}b). 
In practice, however, 
multiple elements in the growth process can act as a buffer. For example, 
when As is provided to the growth chamber, the chamber itself and the
gold droplet used as a growth catalyst still contain excess P atoms. 
The excess P content can be expected to decay exponentially with some
delay length $l$ as P and As atoms are incorporated into the nanostructure.
Similarly, after switching off the As supply, the excess As content 
decays exponentially. The resulting As concentration as a function of 
the $z$-coordinate is depicted in Fig.~\ref{fig:delay}b. 

The effects of delayed As incorporation in the dot are hard to estimate 
in advance because the quantum dot region is no longer clearly defined.
The volume with non-zero As content increases but the strength of the 
confining potential decreases. 
Furthermore, strain effects might obscure the picture.
Therefore, numerical studies of the effects of delayed As incorporation
are necessary.

Figures ~\ref{fig:delay}a and c depict the 
single-particle states as well as the Coulomb renormalization of the 
lowest-energetic manybody complexes as a function of the delay length $l$.
It is found that with increasing delay length the lowest conduction band levels
shift upward while the highest valence band states shift downwards.
This indicates that electrons and holes tend to be more confined to
the region with maximal As content in structures with delayed As incorporation
compared with the situation in dots with a uniform distribution of the same
number of As atoms. The manybody energy renormalization is only marginally 
affected.

\subsection{Convergence\label{sec:convergence}}
\begin{figure}
\includegraphics[width=0.95\linewidth]{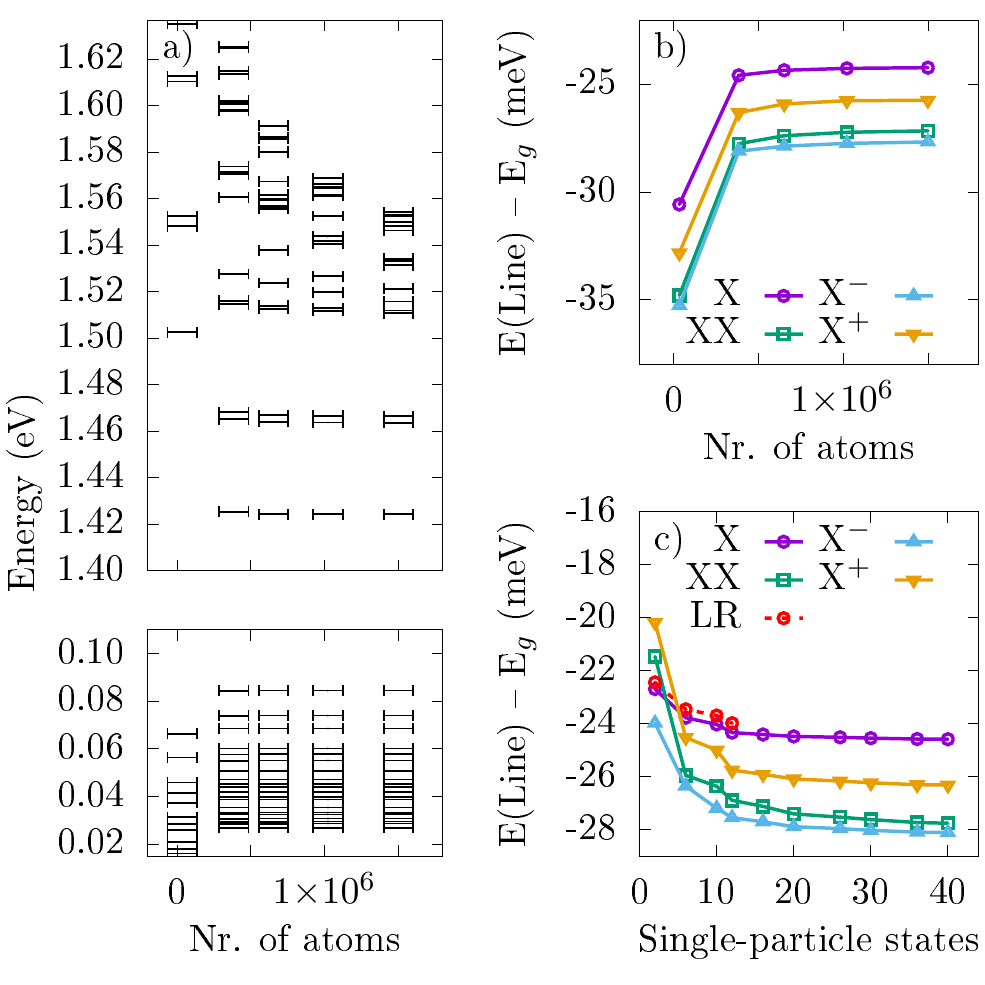}
\caption{\label{fig:convergence} Conduction and 
valence band states (a) as well as the lowest-energetic exciton
biexciton and trion lines (b) as a function of  
the number of atoms in the computational box. 
(c): Evolution of the positions of spectral lines as a function of
the number of single-particle states taken into account in the basis for
the configuration-interaction caclulation. 
}
\end{figure}

For the simulation of single-particle states 
one has to define a computational box with a finite size.
For excitonic complexes, the configuration-interaction calculation 
requires a truncation of the number of single-particle states from 
which the manybody Hilbert space is constructed.
Both convergence parameters, the volume or number of atoms in the 
computational box and the number of single-particle states entering the
manybody calculation, might have an impact on the accuracy of the calculations
and limit the size of the systems that can be investigated using 
our numerical toolkit.

Figure~\ref{fig:convergence}a shows the energies of the 40 lowest conduction 
and valence band states as a function of the number of atoms
in the computational box up to 1.5 million atoms
for the same quantum dot as discussed in Fig.~\ref{fig:SP}. 
The first computational box with about 40,000 atoms consists of only the 
quantum dot itself. It can be clearly seen that this computational box
is not sufficient for an accurate description, because
the wave functions significantly leak outside of the quantum dot region.
After a box size with about 400,000 atoms, 
the lowest few energy levels in the conduction band as well as
all considered valence band states remain virtually unchanged.
However, new levels appear higher in the conduction band around the
first d-levels.
This is due to the fact that, because of the band alignment 
between InP and InAs,
conduction band electrons experience only a shallow confining potential.
The new levels correspond to deconfined states and indicate the onset of
a continuum of states in the surrounding InP nanowire.
However, this has only a marginal quantitative effect on the positions
of the spectral lines of excitonic complexes, which are depicted in 
Fig.~\ref{fig:convergence}b, because the charge density of the deconfined
states outside of the dot are spatially separated from the relevant 
states within the dot.

The convergence of the configuration-interaction calculation as a function of 
the number of single-particle states taken into account in the basis of 
Slater determinants is shown in Fig.~\ref{fig:convergence}c. Note that, because
the onsite terms are numerically more demanding, we calculate only Coulomb 
matrix elements between the 12 single-particle states per band (electrons or
holes) for the onsite terms and add them to the long-range Coulomb terms, 
which we calculate for up to 40 single-particle states per band.
For such large structures as the systems under consideration here, the 
onsite terms have a marginal effect, as can be seen by the positions of the
excitons lines denoted as LR and depticed in red 
in Fig.~\ref{fig:convergence}c, where we only take long-range Coulomb matrix 
elements into account for the manybody calculation and drop the onsite terms.

The convergence studies suggest that for the discussion of a single 
InAsP quantum dot a computational box size of about 400,000 atoms is 
sufficient and convergence of the manybody problem for the lowest-energetic
excitonic complexes is reached when 20 single-particle states per band are 
taken into account.

\section{Conclusion}

We have presented an atomistic theory of the electronic and optical properties of hexagonal 
InAsP quantum dots in InP nanowires in the wurtzite phase.

We have obtained tight-binding parameters 
for atomistic simulations of InAs$_x$P$_{1-x}$ nanowire quantum dots 
in the wurtzite phase by \emph{ab initio} methods.
Using a new, highly parallelized code, we have performed calculations of
single-particle states as well as spectra of excitonic complexes for 
quantum dots with varying sizes and As concentrations as well as different
random distributions of As atoms within the dot.

The low energy single-particle states, in particular the conduction band
states, form a shell structure and show highly symmetric wave functions 
that can be classified in terms of s-, p- or higher states in close analogy to
flat cylindrical quantum dots.
While the growth of quantum dots in nanowires allows a fabrication of hexagonal
dots with predefined sizes, some residual random fluctuations of the electronic 
and optical properties, such as the positions of the spectral line in 
the range of a few meV, are found as a result of the random alloying of 
As atoms in the InP matrix. 

Varying the quantum dot size reproduces characteristic confinement effects, 
where, e. g., the single-particle gap decreases for larger dots. Similarly,
increasing the As/P ratio within the dot leads to deeper confining 
potentials for electrons and hole and, thus, to a reduced single-particle gap.
Furthermore, when the distribution of As atoms along the growth direction 
is smeared out because of a delayed incorporation of As atoms during the
growth process, we find the gap to increase.

The simulation of excitonic complexes such as excitons, biexcitions, and trions
consistently predicts spectral lines 
where the biexciton binding energy is positive and varies with a 
stardard deviation of 0.7 meV about the mean value of 2.4 meV, 
the negative trion is lower in
energy than the biexciton line and positive trion is higher in energy compared
to the biexciton, while the positive trion is close to the bright exciton line.
Varying the parameters of the quantum dot has only a minor influence on the
Coulomb renormalization of the energies of manybody complexes, but 
affects excitons, biexcitons and trions in a similar way, so that we cannot
identify an unambiguous fingerprint 
of the geometrical properties on, e.g., the relative
positions of the spectral lines originating from different complexes.

For hexagonal or disk-shaped [111] grown nanowire quantum dots, 
the fine structure splitting between the bright exciton states 
has been predicted to vanish on grounds of symmetry in 
Ref.~\onlinecite{Bester_smallFSS}. 
Our calculations predict an average fine structure splitting of 7.1 $\mu$eV 
with a standard deviation of 4.6 $\mu$eV. This is in line with 
atomistic tight-binding calculations for cylindrical InAsP quantum dots 
using zincblende parameters\cite{ZielinskiFSS}.
In Ref.~\onlinecite{Dalacu_entangled_2014} the fine structure splitting 
for a number of nanowire quantum dots has obtained experimentally yielding 
a mean of 3.4~$\mu$eV with a standard deviation of 3.0~$\mu$eV.

An experimental spectra of a wurtzite InAsP nanowire quantum dot is
presented in Fig. 1 of Ref. \onlinecite{Dalacu_entangled_2019}.
The exciton line is located at 1.388 eV. The biexciton binding energy 
is positive and around 2.2 meV. The negative trion line is found to be lower
in energy than the biexciton line while the positive trion is slightly above
the exciton line. The fine structure splitting for the dot in 
Ref. \onlinecite{Dalacu_entangled_2019} is 3.3~$\mu$eV.
Thus, the relative positions of the spectral lines for this dot agrees well 
with our predictions. 

However, some nanowire quantum dots have also been reported\cite{Dalacu_pureWZ}
to show negative biexciton binding energies of about -1.5 meV and a number
of effects have not yet been accounted for in the modelling. For example,
studies of laser-induced atom intermixing\cite{Dalacu_intermixing} 
highlight the possibility of diffusion processes and suggests the existence
of some defects. Furthermore, as in Stranski-Krastanov-grown quantum dots,
there may be a tendency for As atoms to cluster so that the distribution
is no longer uniform within one monolayer.

The tight-binding parameters obtained here pave the way for further studies 
in this direction. Furthermore, our convergence studies have proven the
feasibility of simulations with more than one million atoms, which is 
sufficient to investigate nanowires with two or more quantum dots.

\acknowledgements
M.C. gratefully acknowledges funding from the Alexander-von-Humboldt 
foundation through a Feodor-Lynen research fellowship. 
P.H. and M.C. thank NSERC QC2DM Project and uOttawa Chair in Quantum Theory of Materials, Nanostructures and Devices for support. 
P.H. and M.C. acknowledge computational resources 
provided by Compute Canada.
Discussions with D. Dalacu, P. Poole and D. Gershoni are acknowledged. 

\bibliography{Bibliography}
\end{document}